\documentclass[conference]{IEEEtran}
\usepackage[noadjust]{cite}

\usepackage{caption}
\usepackage{array}
\usepackage{mathtools}
\usepackage{graphicx}
\usepackage{algorithm}
\usepackage{algpseudocode}
\usepackage{subcaption}
\usepackage{subfig}
\usepackage{multirow}
\usepackage{array}
\usepackage{color}
\usepackage[T1]{fontenc}
\usepackage[scaled]{helvet}
\usepackage{luximono}

\usepackage{tablefootnote}

\ifCLASSINFOpdf
\else
\fi
\begin{document}
\bstctlcite{IEEEexample:BSTcontrol}
%
\title{Throughput and Range Characterization of IEEE 802.11ah}

\author{
	\IEEEauthorblockN{Victor Ba\~{n}os-Gonzalez\IEEEauthorrefmark{1}, M. Shahwaiz Afaqui\IEEEauthorrefmark{1}, Elena Lopez-Aguilera\IEEEauthorrefmark{1}, Eduard Garcia-Villegas\IEEEauthorrefmark{1}}
	\IEEEauthorblockA{\IEEEauthorrefmark{1} UPC-BarcelonaTECH
	}}


%


\maketitle

\begin{abstract}
The most essential part of Internet of Things (IoT) infrastructure is the wireless communication system that acts as a bridge for the delivery of data and control messages. However, the existing wireless technologies lack the ability to support a huge amount of data exchange from many battery driven devices spread over a wide area. In order to support the IoT paradigm, the IEEE 802.11 standard committee is in process of introducing a new standard, called IEEE 802.11ah. This is one of the most promising and appealing standards, which aims to bridge the gap between traditional mobile networks and the demands of the IoT. In this paper, we first discuss the main PHY and MAC layer amendments proposed for IEEE 802.11ah. Furthermore, we investigate the operability of IEEE 802.11ah as a backhaul link to connect devices over a long range. Additionally, we compare the aforementioned standard with previous notable IEEE 802.11 amendments (i.e. IEEE 802.11n and IEEE 802.11ac) in terms of throughput (with and without frame aggregation) by utilizing the most robust modulation schemes.  The results show an improved performance of IEEE 802.11ah (in terms of power received at long range while experiencing different packet error rates) as compared to previous IEEE 802.11 standards.

\end{abstract}


%
\IEEEpeerreviewmaketitle

\section{Introduction}
\label{Introduction}
\indent The key to the concept and success of smart cities (an application area of IoT paradigm), that aims to improve the quality of life and alleviate public services in urban centers, is particularly based on exponential growth of different radio technologies. Smart cities take advantage of communication networks and sensors (i.e. IoT devices) to optimize various logistical operations (e.g. transport, electrical, etc.) to improve the quality of life of people residing in the cities. In today's smart cities, cellular and wireless sensor networks are the dominant technologies used to relay information towards a central processing office. Since the amount of information generated over such scenario is assumed to be huge and increasing (due to the increase of connected devices), there is a need to adopt universally accepted, cost effective and scalable communication technology within IoT framework. IEEE based WLANs (due to their ease of deployment and cost efficiency) could be used as viable alternative technology for smart cities only if the limitations of high power consumption and limited number of associated stations are overcome.\\ 
\indent In recent years, tremendous proliferation of IEEE 802.11-based WLAN has been witnessed. The wider acceptance of IEEE 802.11 has resulted in mass deployments in diverse environments (e.g. homes, offices, streets, campuses, etc.) where different devices (e.g. smartphones, laptops, tablets, wearables, etc.) utilize the aforementioned standard as a major access method to connect to the Internet. The IEEE 802.11 in its current version 802.11 a/g, n or ac was not focused on developing any IoT specification. Actually, a task group of IEEE 802.11 (called TGah) is working on the draft version~\cite{stdah} of a new standard, the IEEE 802.11ah, the 802.11 approach to the IoT. This standard is intended to provide low cost mode of operation, with greater coverage area, and thousands of associated stations per cell. To assess how this new IEEE 802.11ah standard adds value to the 802.11 family in terms of range and throughput for new use cases, we compare its performance against current IEEE 802.11 amendments.
\vspace{-0.5em}
\subsection{Related Work}
IEEE 802.11ah standard aims to organize communication between various devices used in IoT applications such as smart grids, smart meters, smart houses, healthcare systems and smart industry. In order to expose the key mechanisms of the upcoming IEEE 802.11ah amendment, the authors in \cite{Khorov201553} provide a comprehensive overview. Similarly, in \cite{7060496} and \cite{7000982} authors detail the distinct features of IEEE 802.11ah. In \cite{6846178}, the authors highlight the importance of IEEE 802.11ah standard as one of the key enabling technology for low cost, energy efficient and massive deployment for IoT devices in future. Furthermore, the authors evaluate maximum achieved throughput in three different Modulation and Coding Schemes (MCS) of IEEE 802.11ah using significant assumptions. Also in \cite{7060496}, the authors show performance results for measurement data of IEEE 802.11ah in terms of rate and range. They compare IEEE 802.11ah with 802.11b and 802.11n for three indoor cases without taking into account outdoor scenario that is the most usable case for IEEE 802.11ah, (in our current work we include comparison between indoor and outdoor use cases). The work in \cite{7101216} provides a comprehensive overview of IEEE 802.11ah. Furthermore, the authors summarize standardization procedures as well as the technical challenges expected in the adaptation of IEEE 802.11ah standard. In \cite{ahusecases} , the authors define different innovative use cases for IEEE 802.11ah standard. Among the proposed use cases, the authors highlight an interesting case where IEEE 802.11ah standard will be able to provide appropriate feature as a backhaul link to accommodate traffic exchange over long distances (i.e. leaf sensors and stream of camera images or surveillance videos).

Our contributions in this paper are, in the first place, a brief but updated overview of the IEEE 802.11ah standard, according to the latest version of the document. Second, we perform a novel performance comparison of IEEE 802.11ah standard with previous amendments (i.e. IEEE 802.11a/n/ac) in terms of range and throughput (with and without frame aggregation). By doing so, we highlight the significance of IEEE 802.11ah as one of the most effective technologies to provide good throughput at larger distances and thus give substance to IEEE 802.11ah backhaul use case. \\
\indent The remainder of the paper is organized as follows. In Section II we present an overview of IEEE 802.11ah technology.  Section III gives the details of evaluation scenario and includes an analytical model for throughput analysis. In Section IV the performance evaluation in terms of coverage range and throughput is exposed. Finally, Section V concludes the paper and presents future work.
\vspace{-.2em}

\section{Notable PHY and MAC features of IEEE 802.11ah}
\label{Notable PHY and MAC features of IEEE 802.11ah}
This standard intends to modify the current IEEE 802.11 standard (at PHY and MAC layer) in order to extend it to operate below 1GHz (S1G) for ubiquitous access in less interfered frequency band and to support large number of associated stations within the network. Due to the deficiency encountered in the scarce availability of sub 1 GHz bands, the physical layer modifications are intended to improve the spectral efficiency. Furthermore, due to the intention of having numerous IoT devices contending for the shared resources, the MAC of the new amendment is designed to administer scalable operation. In addition, the proposed MAC features assist to improve power efficiency among stations that have limited energy resources. Smart grids, home automation, smart cities and smart health applications typically require 100Bytes of data size, coverage ranges up to 1km, more than 1000 devices connected per access point (AP) and an average data rate of 100kbps. It is pertinent to mention here that, due to the redesigned MAC and PHY layer, the new standard is not anticipated to be backward compatible. Table \ref{tab:1}  summarizes the key features of IEEE 802.11ah and compares them with previous proposed amendments of IEEE 802.11.
 \begin{table}[h]
 	\vspace{-.5em}
\caption{Comparison of IEEE 802.11 standards.}
\label{tab:1}
\centering
\scalebox{0.9}{
\begin{tabular}{|>{\centering\arraybackslash}p{2cm}|>{\centering\arraybackslash}p{1.25cm}|>{\centering\arraybackslash}p{1.25cm}|>{\centering\arraybackslash}p{1.25cm}|>{\centering\arraybackslash}p{1.25cm}|}
\hline
\bf{}&\bf{802.11a/g}&\bf{802.11n}&\bf{802.11ac}&\bf{802.11ah}\\ \hline
\bf{Antenna Configuration}&	$1\times1$ SISO & $4\times4$ MIMO &	$8\times8$ MIMO & $4\times4$ MIMO \\ \hline
\bf{Highest Order Modulation}	& BPSK to 64-QAM &	BPSK to 64-QAM	& BPSK to 256-QAM &	BPSK to 256-QAM \\ \hline
\bf{Channel Bandwidth}	& 5, 10 MHz (11a), 20 MHz (11a/g)	& 20 and 40 MHz mode &	20, 40, 80 and 160 MHz &	1, 2, 4, 8, and 16 MHz \\ \hline
\bf{FFT Size} &	64 &	64 (20 MHz), 128 (40MHz) &	64, 128, 256 and 512 &	32, 64, 128, 256 and 512 \\ \hline
\bf{Year Approved} &	1999/2003 &	2009 &	2014 &	2016 (draft) \\ \hline
\bf{Min. and Max. Bit rate} &	6 and 54 Mbps &	6.5 and 600 Mbps &	6.5 and 6933.3 Mbps &	0.15 and 347 Mbps \\ \hline
\bf {Max. num. of supported STAs} &	2007 &	2007 &	2007 &	About 8000 \\ \hline
\end{tabular}}
\end{table}
\\In the following section, we give a brief description of PHY and MAC layer enhancements proposed for IEEE 802.11ah standard.
\vspace{-.3em} 
\subsection{PHY layer}
The physical layer of IEEE 802.11ah inherits its main characteristics from IEEE 802.11ac, but is adapted to operate at S1G frequency band. It is designed to operate by utilizing Orthogonal Frequency Division Multiplexing (OFDM) along with Multiple Input Multiple Output (MIMO) including Multi-user MIMO (MU-MIMO) over the downlink. Additionally, it supports various MCSs (i.e. from MCS0 to MCS10). However, given limited capabilities and limited data transfer requirement for certain applications, high-order modulations or even multiple streams are not likely to be widely supported or required for first Wi-Fi certifications. Table \ref{tab:2} highlights the key PHY layer characteristics of 802.11ah. 
\begin{table}[h]
\caption{PHY layer parameters for IEEE 802.11ah.}
\label{tab:2}
\centering
\scalebox{0.9}{
\begin{tabular}{|>{\centering\arraybackslash}p{2.5cm}|>{\centering\arraybackslash}p{2cm}|>{\centering\arraybackslash}p{2cm}|>{\centering\arraybackslash}p{1.5cm}|}
\hline
\bf{Parameter}&\bf{Values}&\bf{Parameter}&\bf{Values}\\ \hline
\bf{Carrier Frequency} &	863-868MHz Europe
 902-928MHz US	& \bf{Bandwidth (MHz)} &	1, 2, 4, 8, 16 \\ \hline
 \bf {Number of data/total subcarriers per OFDM symbol} &	24/32 (1MHz)
 52/64 (2MHz)
 108/124 (4MHz)
 234/256 (8MHz)
 468/512 (16MHz)	&
 \bf{Preamble type}	& Short (1MHz). Long (2,4,8,16MHz) \\ \hline
\bf{ Number of spatial streams (SS)} &	$1-4$ &	\bf{Subcarrier spacing}	& 31.25 (kHz) \\ \hline
\end{tabular}}
\vspace{-1.8em}
\end{table}
In the following section, we expose the main physical layer amendments proposed for IEEE 802.11ah that substantiates its operation for IoT devices.
\subsubsection{Available spectrum}
Due to limited availability of license exempt spectrum in 1GHz and owing to the intention of enabling Wi-Fi devices to gain access of channel for short-term transmissions, the basic channel width utilized in IEEE 802.11ah is 1MHz but channel bonding can be applied to create up to 16MHz-wide channels (cf. Tables I and II). However, it is expected that early commercial devices support up to 4MHz.
\subsubsection{Transmission modes}
The main requirement for this amendment is to extend the range of operation and thus to facilitate IoT devices (placed at greater distances) that require low data rates. This aforementioned requirement is fulfilled by introducing 1MHz-wide transmissions and by using a new MCS index (called MCS10). This scheme is effectively MCS0 with an addition of 2x repetition (where OFDM symbols repetition is performed with subcarrier permutation). Apart from 1MHz, IEEE 802.11ah standard also supports 2, 4, 8 and 16MHz where the PHY layer is effectively 10 times down-clocked version of IEEE 802.11ac, i.e. OFDM symbol in IEEE 802.11ah standard is 10 times longer than IEEE 802.11ac.
\subsection{MAC layer}
The MAC layer of IEEE 802.11ah includes improvements to specifically address the requirements of long range communication and IoT use cases. Furthermore, the MAC layer is optimized to encompass low power mode of operation and methods to support large number of devices over a single cell. In the following section, we describe in detail the MAC layer enhancements proposed by the IEEE 802.11ah.
\subsubsection{Compact frame format to increase throughput}
IEEE 802.11ah stations in most of the use cases are expected to operate at low data rates and intend to exchange small data frames. Specifically for IoT devices, the overhead associated with frame headers (e.g. MAC header) may be considerable when compared to the size of the payload.  In order to counter overheads and to increase the efficiency and thus, overall throughput, the MAC design of IEEE 802.11ah introduces compact frame formats.

\hspace{-1em} {a. Short MAC header format}
\vspace{0.5em}
\\
 \hspace{1em}The significant change in the new header design is the inclusion of only two mandatory address fields as compared to four addresses fields present in the legacy MAC header. The Quality of Service (QoS) and High Throughput (HT) fields are shifted into SIG field in PHY header and Duration/ID field is removed (because the virtual carrier sensing is not used while utilizing short MAC header). Thus, the short MAC header is able to reduce the overhead (from 30Bytes to 18Bytes).

\hspace{-1em} {b. Short MAC control frames}
\vspace{0.5em}
\\
 \hspace{1em}To reduce the overhead induced by control frames, the IEEE 802.11ah utilizes Null Data Packets (NDP), which contain PHY header without any data. Different control frames (e.g. CTS, ACK, PS-Poll frame, etc.) are substituted by NDP frames to reduce protocol overhead. 
\subsubsection{Restricting the effects of fading}
In order to tackle time and frequency selective fading over narrow band channels, the IEEE 802.11ah implements a new feature called Sub-Channel Selective Transmission (SST). This scheme allows stations to rapidly switch among specific set of sub-channels during transmission where the channel is selected based on measurements indicating short term fading conditions and/or the level of interference from other stations.
\subsubsection{Large number of stations with Hierarchical Grouping}
For increasing the number of supported stations, IEEE 802.11ah utilizes a novel hierarchical Association Identifier (AID) structure. The AID assigned by the AP during association consists of 13bits and thus the number of stations that it can associate is up to $2^{13}-1$ (8,191). AID structure consists of four hierarchical levels (i.e. page, block, sub-block, and station's index in sub-block). IEEE 802.11ah utilizes the aforementioned structure to group stations based on similar characteristics (e.g. traffic pattern, location, battery level etc.).  
\subsubsection{Channel access}
The IEEE 802.11ah defines a new contention-free channel access period called Restricted Access Window (RAW). This access method is designed to reduce collisions by improving the channel efficiency. The AP coordinates the uplink channel access of the stations by defining RAW time intervals in which specific class of devices are given exclusive access of the shared medium.
\subsubsection{Power saving mode}  
In order to support numerous IoT devices, the TGah has placed paramount importance on developing and enhancing power saving mechanisms. IEEE 802.11ah proposes to use \textit{speed frame exchange} method that enables an AP and non-AP station to exchange a sequence of uplink and downlink frames during a reserved Transmit Opportunity (TXOP). This scheme helps to extend battery life of stations by keeping them awake for shorter duration of time. Instead of using same \textit{Max idle period\footnote{Time during which a non AP station can refrain from transmitting to the AP before being disassociated due to inactivity}}  for all nodes (i.e. 18.64hrs), IEEE 802.11ah aims to utilize different periods for different devices (i.e. from 18.64hrs to 186hrs). Furthermore, IEEE 802.11ah enables a station to inform the AP about the duration of time it intends to remain in sleep mode. During the sleep mode, the station is not intended to listen to beacons and then it is able to reduce its power consumption.
\vspace{-.3em} 
\subsection{Use Cases}
IEEE 802.11ah presents three basic use cases \cite{Khorov201553}:
\subsubsection{Smart Sensor and meters}
In this use case, the AP covers a high number of sensor devices. There are thousands of stations contending for the channel, operating at long transmission ranges along with stationary mobility. The AP to station ratio is of 1/6000. The most common scenarios are large indoor spaces and outdoor in urban, suburban and rural environments. In these scenarios, devices typically send traffic of the order of 100kbps of bit rate, consisting of short frames. 
\subsubsection{Backhaul aggregation and extended range hotspot}
IEEE 802.15.4 sensor devices show extended battery life, however, the transmission range and available data rates are very low (some kbps). Thus, a scenario in which IEEE 802.15.4 routers gather data from leaf devices (i.e. sensors) and forward information to servers using IEEE 802.11ah links results very attractive (cf. Figure 1). This use case is addressed to outdoor industrial and rural environments with lower than 1Mbps of bit rate per station, along with stationary or low mobility devices. The AP to station ratio is of 10/500.
\subsubsection{Extended range hotspot and cellular offloading}
Both high throughput and long transmission range make S1G attractive for extending hotspot range and for traffic offloading in mobile networks, which is a significant issue for operators and vendors due to mobile traffic explosion. IEEE 802.11ah will provide real additional value, especially in countries with wide available S1G spectrum (e.g. USA). This use case is addressed to outdoor use in urban and suburban environments with less than 20Mbps of bit rate, along with pedestrian mobility. TGah shall consider traffic models for TGah-specific applications such as: web browsing with 256kbps per link and a MSDU size of 1000Bytes on TCP, video/audio streaming with 100kbps to 4Mbps per link and a MSDU size of 512Bytes on UDP, and audio streaming with 64kbps to 256kbps per link and a MSDU size of 418Bytes (UDP). The AP to station ratio is of 1/50.
\begin{figure}[htb]
\centering
\graphicspath{/AHlatex/Figures}
\includegraphics[width=0.48\textwidth]{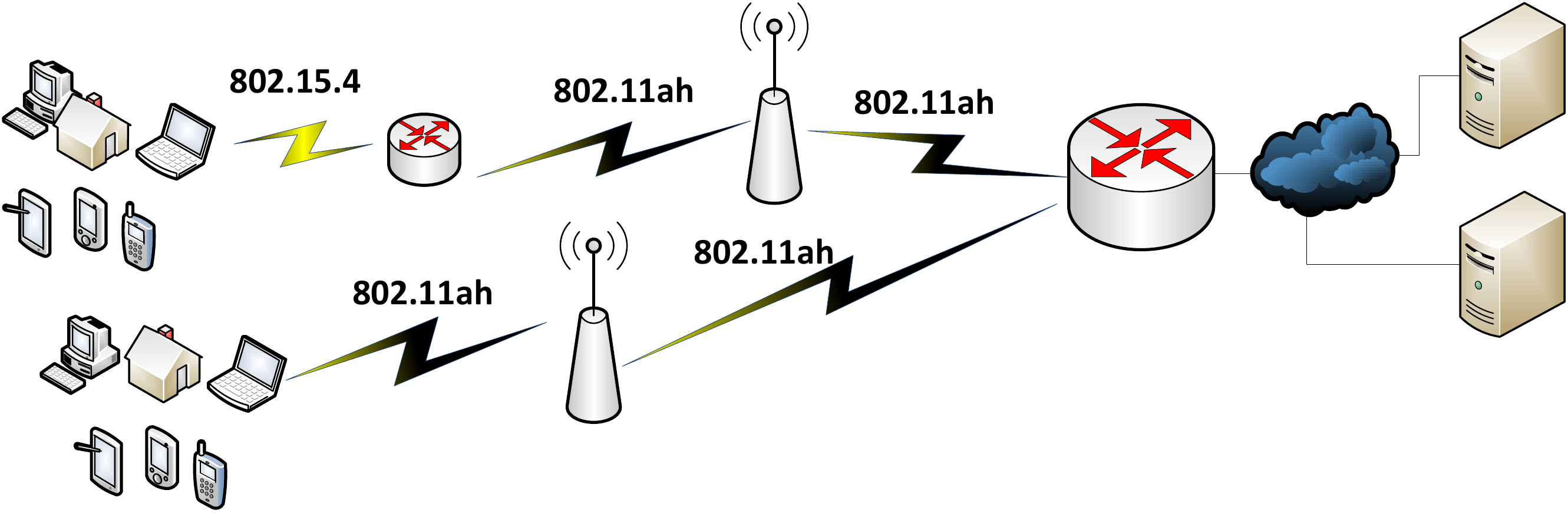}\\
\caption{Backhaul Sensor Network.}
\label{fig:fig1}
\vspace{-.5em}
\end{figure}
\vspace{-.5em}
\section{Evaluation Scenario}
\label{Evaluation Scenario}
We present a comparison between IEEE 802.11ah and 802.11(a, n, ac) amendments in terms of throughput and transmission range. Our evaluation scenario consists in a single radio link composed of two stations (a transmitter and a receiver) that exchange data frames with different payload sizes. Section IV shows in detail the configuration used at stations. The most robust MCS is considered for the different IEEE 802.11 specifications. Specifically, with regard to IEEE 802.11ah, MCS10 is taken into account (i.e. 1MHz channel bandwidth, 1 spatial stream, long guard interval and 150Kbps of bit rate). Concerning IEEE 802.11(n and ac), MCS0 with 20MHz channel bandwidth, 1 spatial stream, long guard interval and 6.5Mbps is considered. Table III shows PHY and MAC parameters employed.
\vspace{-0.5em}
\subsection{Channel model}
We consider the path loss models chosen by TGah \cite{channelmodel}:
Macro deployment: outdoor scenario with antenna height of 15m above rooftop:
 \begin{equation}
\label{eq:eq1}
PL= 8 + 37.6log_{10}(d)
\end{equation}
\indent where \textit{d} corresponds to the distance in meters between transmitter and receiver, and radio frequency carrier is 900MHz. For other frequencies, a correction factor of $21·log_{10}$ (f/900MHz) should be applied.
Pico/hot zone deployment:  outdoor scenario with antenna placed at rooftop:
 \begin{equation}
\label{eq:eq2}
PL= 23.3 + 36.7log_{10}(d)
\end{equation}
\indent where the same conditions regarding distance and frequency as in equation. (\ref{eq:eq1}) are applied. TGah indoor path loss model: it is modelled by directly scaling down the frequency operations of the TGn path loss model. It consists of the free space loss model (slope of 2) up to a breakpoint distance (\textit{$d_{BP}$}), and employs a slope of 3.5 after the breakpoint. We consider the large indoor open space scenario with non-line-of-sight (NLoS) conditions (Model C with \textit{$d_{BP}$} of 5m), and with line-of-sight (LoS) conditions (Model D with \textit{$d_{BP}$} of 10m). Both indoor channel models would correspond to a factory/warehouse type of environment\footnote{Note that TGah indoor channel propagation loss model was recently amended according to \cite{CHmodel}}.
\begin{equation}
\label{eq:eq3}
L(d)=\begin{cases}
L_{FS}(d)=20log_{10}\left(\frac{4\pi f_c}{c}\right),& \text{if } d\leq d_{BP}\\
L_{FS}(d_{BP}) + 35log_{10}\left(\frac{d}{d_{BP}}\right),& \text{if } d> d_{BP}\\
\end{cases}
\end{equation}
\indent where \textit{d} corresponds to the distance in meters between transmitter and receiver, \textit{fc} is the center carrier frequency in MHz and \textit{c} the speed of light in m/s.
\newcolumntype{L}{>{\centering\arraybackslash}m{1.1cm}}
\begin{table}[h]
	\vspace{-1.5em}
\caption{MAC/PHY Parameters.}
\label{tab:3}
\centering
\normalsize
\scalebox{0.55}{
\begin{tabular}{|>{\centering\arraybackslash}p{1.5cm}|L|L|L|L|L|L|L|L|L|}
\hline
\parbox[c][5ex]{5ex}{\centering}
\bf{Specifications}
&\bf{SIFS $\mu$s}&\bf{DIFS $\mu$s}&\bf{Tpream
-ble and header $\mu$s}&\bf{MAC and LLC Header Size Bytes}&\bf{Signal Extensions $\mu$s}&\bf{TSym $\mu$s}&\bf{TSlot $\mu$s}&\bf{CWmin}&\bf{CWmax}\\ \hline
\parbox[c][5ex]{5ex}{\centering}
\bf{802.11ah Short MAC Header}& 160 & 264 & 560 & 26 & n/a & 40 & 52& 15 & 1023\\ \hline
\parbox[c][5ex]{5ex}{\centering}
\bf{802.11ah long MAC Header}& 160 & 264 & 560 & 36 & n/a & 40 & 52& 15 & 1023\\ \hline
\parbox[c][5ex]{5ex}{\centering}
\bf{802.11ac}& 16 & 34 & 40 & 36 & n/a & 4 & 9 & 15 & 1023\\ \hline
\parbox[c][5ex]{5ex}{\centering}
\bf{802.11n 2.4GHz}& 10 & 28 & 36 & 36 & 6 & 4 & 9 & 15 & 1023\\ \hline
\parbox[c][5ex]{5ex}{\centering}
\bf{802.11n 5GHz}& 16 & 34 & 36 & 36 & 0 & 4 & 9 & 15 & 1023\\ \hline
\parbox[c][5ex]{5ex}{\centering}
\bf{802.11a}& 16 & 34 & 20 & 36 & n/a & 4 & 9 & 15 & 1023\\ \hline
\end{tabular}}
\vspace{-1.5em}
\end{table}
\subsection{Throughput model}
Initially, we consider ideal transmission conditions. Then, the throughput expression \textit{S} in Mbps is as follows \cite{ETT:ETT1335}:
\begin{equation}
\label{eq:eq5}
S= \frac{L_{data}\times8}{T_{message}}
\end{equation}
\indent where \textit{$L_{data}$} corresponds to the payload size and \textit{$T_{message}$} is computed as:
\begin{equation}
\begin{multlined}
\label{eq:eq6}
T_{message}= DIFS + T_{DATA} + SIFS +\\ T_{ACK} + T_{BACKOFF} + 2\delta
\end{multlined}
\end{equation}
\indent DIFS and SIFS are given in Table \ref{tab:3}, $\delta$ is the propagation delay, $T_{ACK}$ corresponds to the duration of an ACK frame and $T_{DATA}$ represents the transmission time of a data frame, which depends mainly on the size of the payload and the PHY rate. $T_{DATA}$ and $T_{ACK}$ computation also depend on the IEEE 802.11 amendment used in the transmission. Under ideal channel conditions we consider $T_{BACKOFF}$ is, on average, $CW_{min}\setminus2$ times $T_{Slot}$. $T_{DATA}$ is computed according to equation (\ref{eq:eq7}). Frame sizes are given in bytes and frame durations in $\mu$s.
\begin{equation}
\label{eq:eq7}
T_{DATA}=  T_{Preamble\&Header}+(T_{Sym}\times N_{Sym})
\end{equation}
\indent where $T_{Preamble\&Header}$ is given in Table \ref{tab:3}, $T_{Sym}$ is the duration of a symbol and $N_{sym}$ is the number of symbols of the PSDU. $N_{sym}$ for the most robust MCS of the different standards is given in equations (\ref{eq:eq8}) and (\ref{eq:eq9}):
\begin{equation}
\label{eq:eq8}
N_{SymAH}= \left\lceil \frac{14+(L_{Header}+L_{data})\times8}{6} \right\rceil 
\end{equation}
\begin{equation}
\label{eq:eq9}
N_{SymAC/N2.4/N5}= \left\lceil \frac{22+(L_{Header}+L_{data})\times8}{26} \right\rceil 
\end{equation}
\indent $N_{symA}$ calculations are the same as in equation (\ref{eq:eq8}) with constant value 22 instead of 14, and $T_{ACK}$ calculation employs previously exposed $T_{DATA}$ equations with 14Bytes instead of $L_{Header} + L_{data}$. Moreover, in case NDP is utilized as an acknowledgement, there is no data field and the number of symbols ($N_{sym}$) is equal to 0. Note that all previous calculations are presented without taking into account reception errors. Subsequently, we consider an error-prone scenario and compute throughput expression (in Mbps) as in equation (\ref{eq:eq5}) just multiplying the numerator per ($1- PER$), where \textit{PER} corresponds to Packet Error Rate and its value depends on the MCS used and the number of transmitted bits. On the other hand, the presence of errors causes retransmissions and therefore, the effect of IEEE 802.11 exponentially increasing backoff mechanism should now be considered as follows \footnote{The number of transmissions is \textit{i}, \textit{PDR(i)} is the probability of a successful reception after \textit{i} transmissions, and $T_{backoff}(i)$ is the average backoff time after \textit{i} consecutive transmissions of the same frame.}:
\begin{equation}
\label{eq:eq10}
T_{BACKOFF}=\sum_{i=1}^{\infty}PDR(i)T_{backoff}(i) 
\end{equation}
Recall that the scenario consists in a radio link with a transmitter and a receiver, i.e. there is no contention with other stations and retransmissions are only due to bit errors caused by noise. A successful transmission requires that neither the data frame nor the ACK are received with errors. Given that the ACK frame is shorter and that it is sent using the most reliable modulation, we consider \textit{PER} corresponding to ACK frames negligible. Then \textit{PDR(i)} is given by:
\begin{equation}
\label{eq:eq11}
PDR(i)=(1-PER)\times PER^{(i-1)}  
\end{equation}
$T_{backoff}(i)$ follows next expression:
\begin{equation}
\label{eq:eq12}
T_{backoff}(i)=
\begin{cases}
\frac{2^{i-1}(CW_{min}+1)-1}{2}\times T_{Slot}  ,& 1\leq i < m \\
\frac{CW_{max}}{2} \times T_{Slot} ,& 1 \geq m\\
\end{cases}
\end{equation}
\indent where \textit{m} is the maximum number of backoff stages and 
corresponds to 6 (i.e. $CW_{max} = 2^6CW_{min}$). $CW_{min}$, $CW_{max}$ and $T_{Slot}$ are all standard-dependent parameters (cf. Table \ref{tab:3}).  For each retransmission, the range of values that can be given to $T_{backoff}(i)$ is doubled until $CW_{max}$ is reached. For A-MPDU calculations, we assume a block ACK (BA) exchange, where the BA frame is always received regardless of the PER. We consider the same error-prone scenario used before, keeping in mind that we are transmitting at the basic and hence the safest rate in all cases. We consider the maximum data frames of size 1500Bytes in IEEE 802.11n/ac. For IEEE 802.11ah, the data frames are limited to 511Bytes at 1 MHz using MCS10 and 1 spatial stream \cite{stdah} (including preamble and header) and thus, the maximum payload is 475Bytes for long header case, and 485Bytes for the short header case. With A-MPDU, throughput is computed according to equation (\ref{eq:eq5}) modifying just the numerator per $(1-PER)L_{data}\times 8\times K$ where K is the number of aggregated frames (of equal size). Note that now the number of symbols $N_{sym}$ is as follows.
\begin{equation}
\resizebox{.95\hsize}{!}{$
N_{SymAH}= \left\lceil \frac{8\times K\times (L_{Header}+L_{data})+14+(K-1)\times(L_{deli}\times 8)}{6} \right\rceil
$}\\
\label{eq:eq13}
\end{equation}
\indent $N_{symAC/N2.4/N5}$ calculations are the same as in equation (\ref{eq:eq13}) with constant value of 22 instead of 14, corresponding to the number of bits in the service field plus the multiplication of the number of tail bits per binary convolutional code (BCC) encoder and the number of BCC encoders. Also the denominator is changed from 6 to 26, which correspond to the number of bits per symbol of the most reliable modulation, $L_{deli}$ is the size of the delimiter between aggregated frames (4Bytes). For $T_{ACK}$ computation, BA frame of 32Bytes is considered.
\vspace{-1em}

\section{Results}
\label{Results}
In this section, we follow the expressions given in Section III to compare the maximum range and expected throughput for different IEEE 802.11 technologies. As mentioned earlier, we first evaluate received power in different propagation scenarios and then we provide throughput estimations as a function of different \textit{PER}. For each of the technologies studied (IEEE 802.11 ah/ac/n/a), we use the PHY configuration providing the longest range; that is, we select the most reliable MCS with the narrowest possible bandwidth (lowest sensitivity required), and we set the maximum transmitted power allowed for each band (cf. Table \ref{tab:4}). Results are given in Table \ref{tab:5}. IEEE 802.11ah benefits from a lower frequency band, incurring in less propagation losses, a narrower bandwidth, improving power spectral density, and a more robust coding scheme. For these reasons, IEEE 802.11ah has the widest coverage.  In all cases, IEEE 802.11ah coverage range shows, at least, a fivefold increase with respect to IEEE 802.11a (second best range), and more than ten times the range provided by IEEE 802.11n in the 2.4GHz band (worst case). None of the indoor scenarios considered in the IEEE 802.11ah use case set (e.g. factory, warehouse, open office, etc.) will require such a long range. In those cases, the increased range offered by IEEE 802.11ah may not compensate for the sacrifice in throughput, as explained next. Note that in  Eq. (\ref{eq:eq3}), is not considered the effect of walls. Also note that we do not consider the gains of the spatial diversity techniques enabled by MIMO technology. Those techniques could increase the range of IEEE 802.11n/ac/ah between 10 and 30m, depending on the scenario and the antenna configuration. Next, in Figures 2 to 4, we show the throughput that could be achieved between a single transmitter and receiver pair at the limits of their coverage. We show throughput values for different payload size (between 12 and 1500Bytes) and different \textit{PER}.
\begin{table}[h]
\caption{PHY configuration for most robust links.}
\vspace{-.1em}
\label{tab:4}
\centering
\scalebox{0.9}{
\begin{tabular}{|>{\centering\arraybackslash}p{2.2cm}|>{\centering\arraybackslash}p{1.25cm}|>{\centering\arraybackslash}p{1.25cm}|>{\centering\arraybackslash}p{1.25cm}|>{\centering\arraybackslash}p{1.25cm}|}
\hline
\bf{Specification}&\bf{802.11ah}&\bf{802.11ac}&\bf{802.11n}&\bf{802.11a\tablefootnote{Note that the IEEE 802.11 standard provides support for the half-clocked and quarter-clocked operation (i.e. 10 and 5MHz) only in Clause 18 (11a).}}\\ \hline
\parbox[c][3ex]{3ex}{\centering}
\bf{Frequency (GHz)} & 0.9& 5.15,5.45 &2.4 & 5.15,5.45 \\ \hline
\parbox[c][3ex]{3ex}{\centering}
\bf{Ptx (mW)} & 1000& 200,1000 & 100 & 200,1000 \\ \hline
\parbox[c][3ex]{3ex}{\centering}
\bf{Sensitivity (dBm)} & -98 & -82 & -82 & -88 \\ \hline
\parbox[c][3ex]{3ex}{\centering}
\bf{Bandwith (MHz)} & 1 & 20 & 20 & 5 \\ \hline
\parbox[c][3ex]{3ex}{\centering}
\bf{Modulation} & BPSK & BPSK & BPSK & BPSK \\ \hline
\parbox[c][3ex]{3ex}{\centering}
\bf{Coding rate} & 1/2 with 2x rep. & 1/2 & 1/2 & 1/2 \\ \hline
\parbox[c][3ex]{3ex}{\centering}
\bf{Bit rate (Mb/s)} & 0.15 & 6.5 & 6.5 & 1.5 \\ \hline
\parbox[c][3ex]{3ex}{\centering}
\bf{Guard interval (ns)} & 8000 & 800 & 800 & 800 \\ \hline
\end{tabular}}
\vspace{-1em}
\end{table}
\begin{table}[h]
	\vspace{-1em}
\caption{Maximum coverage range.}
\label{tab:5}
\centering
\scalebox{0.9}{
\begin{tabular}{|>{\centering\arraybackslash}p{2cm}|>{\centering\arraybackslash}p{1.25cm}|>{\centering\arraybackslash}p{1.25cm}|>{\centering\arraybackslash}p{1.25cm}|>{\centering\arraybackslash}p{1.25cm}|}
\hline
\parbox[c][3ex]{3ex}{\centering}
\bf{Specification}&\bf{802.11ah}&\bf{802.11ac}&\bf{802.11n}&\bf{802.11a}\\ \hline
\parbox[c][3ex]{3ex}{\centering}
\bf{Frequency (GHz)} & 0.9& 5.15,5.45 &2.4 & 5.15,5.45 \\ \hline
\multicolumn{5}{|c|}{\bf{Maximum Distance (meters)}} \\ \hline
\parbox[c][3ex]{3ex}{\centering}
\bf{Macro Deployment} & 1561& 151,221 & 191 & 211,311 \\ \hline
\parbox[c][3ex]{3ex}{\centering}
\bf{Pico-Hot zone} & 721 & 65,93 & 81 & 91,141 \\ \hline
\parbox[c][3ex]{3ex}{\centering}
\bf{Indoor C} & 1138 & 94,142 & 118 & 140,211 \\ \hline
\parbox[c][3ex]{3ex}{\centering}
\bf{Indoor D} & 1531 & 125,191 & 158 & 185,283 \\ \hline
\end{tabular}}
\vspace{-1.5em}
\end{table}
Sensitivity values given in Table \ref{tab:4} guarantee \textit{PER} < 10\%, but a link can be usable at higher error ratios; hence, we provide throughput values for \textit{PER} between 0 and 50\%. Logically, increasing the payload reduces the overhead, enabling higher effective throughput. Note that the use of aggregation would increase IEEE 802.11n/ac/ah efficiency even further, closer to the limit imposed by the PHY rate (cf. Table \ref{tab:4}). However, the impact of aggregation is limited since, using the slowest modulations; we cannot take advantage of a high level of aggregation without exceeding the maximum duration allowed for a frame at the physical layer. Also note that, since the scenarios of interest are outdoors or open indoor spaces, we considered the long guard interval in all cases. 
\begin{figure}[h]
\centering
\graphicspath{/AHlatex/Figures}
\includegraphics[width=0.46\textwidth]{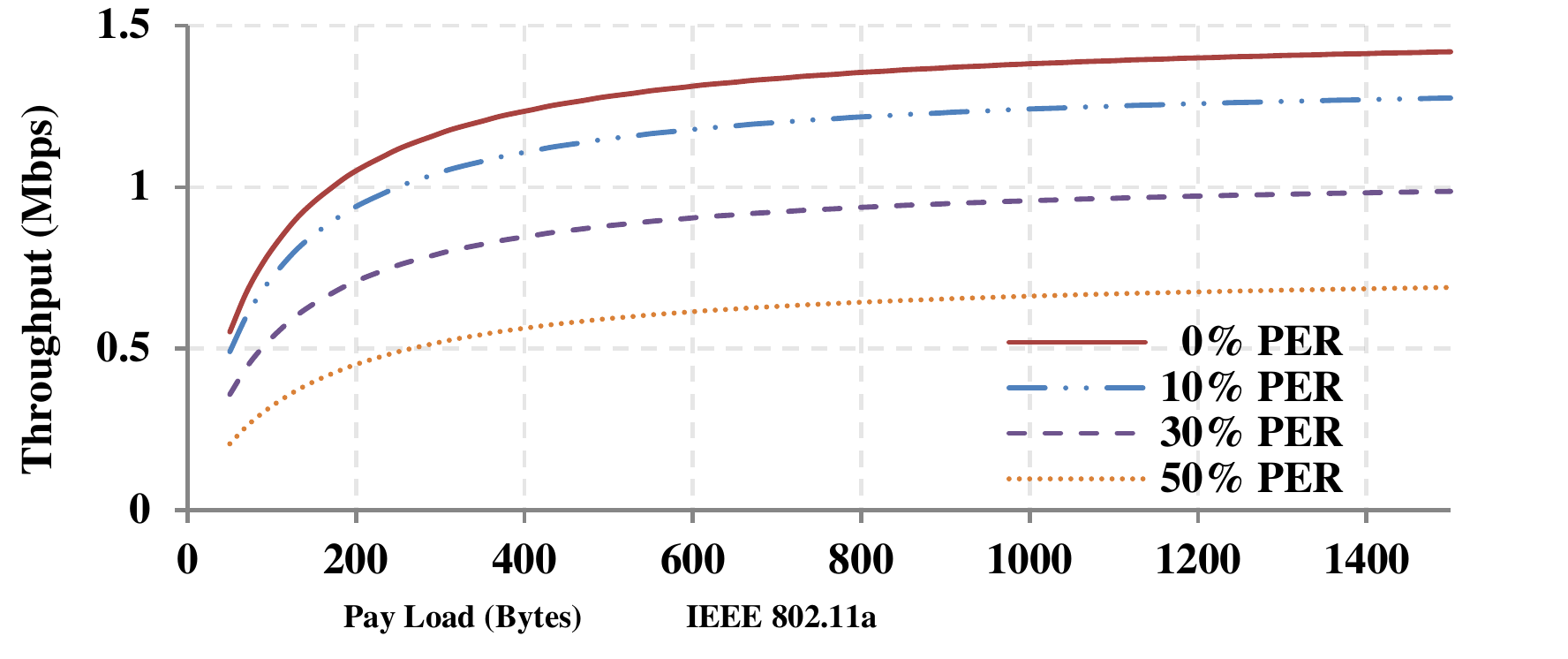}\\
\caption{Throughput vs. Payload size for IEEE 802.11a.}
\label{fig:fig2}
\end{figure}
\begin{figure}[h]
\vspace{-2.5em}
\centering
\graphicspath{/AHlatex/Figures}
\includegraphics[width=0.46\textwidth]{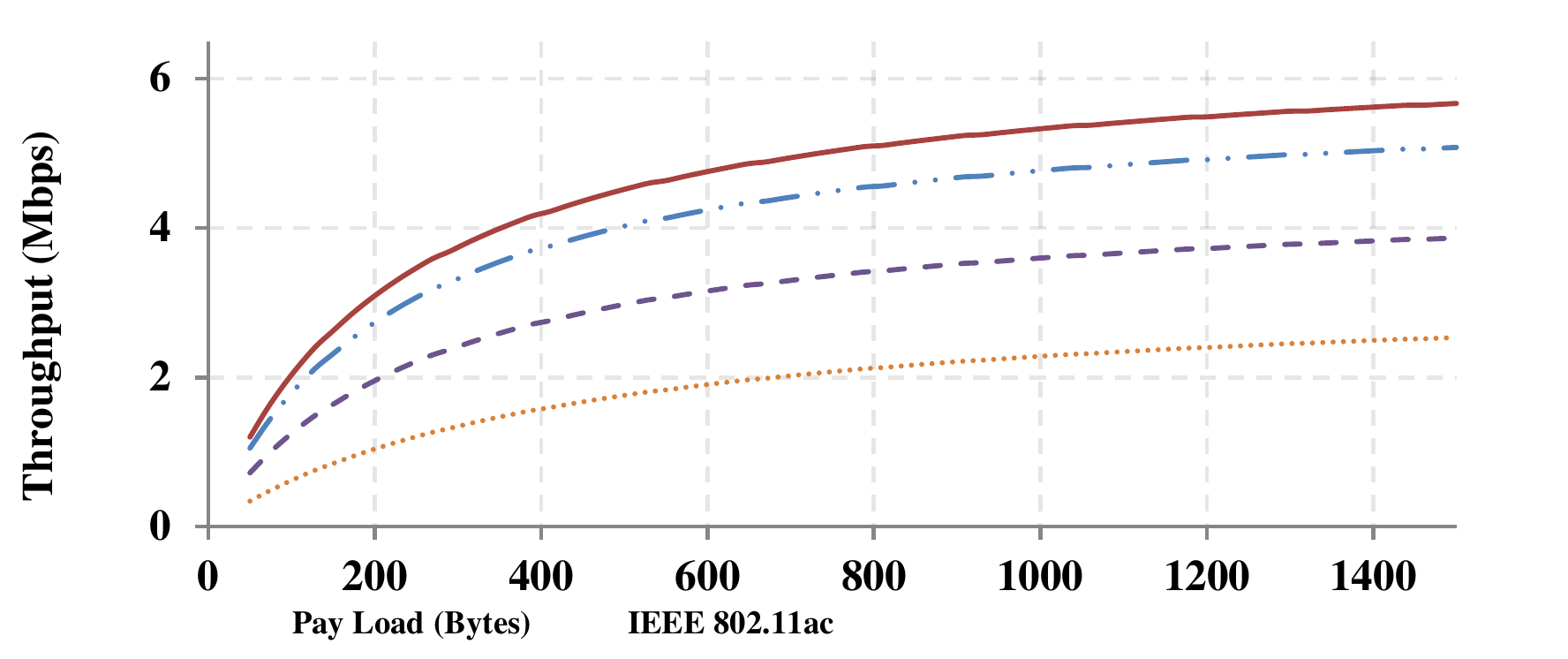}\\
\caption{Throughput vs. Payload size for IEEE 802.11ac.}
\label{fig:fig3}
\vspace{-.5em}
\end{figure}
\\Figure 2 shows the evolution of IEEE 802.11a operating at PHY of 1.5Mbps. It provides a maximum throughput of 1.43Mbps (no errors and payload of 1500Bytes), which is more than halved (0.70Mbps) when \textit{PER} increases to 50\%. With payload of 50Bytes, the maximum throughput is reduced to 0.60Mbps with no errors, and 0.22Mbps with \textit{PER} of 50\%. Due to space constraints and given that the differences between the throughput obtained by IEEE 802.11n and 11ac are minimal at the rate of 6.5Mbps, we only show results for IEEE 802.11ac in Figure 3. For both IEEE 802.11ac and 11n, the maximum throughput with 1500Bytes payload is around 5.6Mbps, which is reduced to 2.5Mbps when the \textit{PER} is 50\%. In the other extreme, i.e. with payload size of 12Bytes, the throughput obtained with \textit{PER} = 50\% is less than one fourth of the throughput with no errors (from 0.33 to 0.08Mbps).IEEE 802.11ah allows a shorter MAC header to reduce overhead. However, we observed minimal throughput improvements (i.e. less than 1\%) when short headers and NDP ACK were used. 
\begin{figure}[h]
\vspace{-.5em}
\centering
\graphicspath{/AHlatex/Figures}
\includegraphics[width=0.48\textwidth]{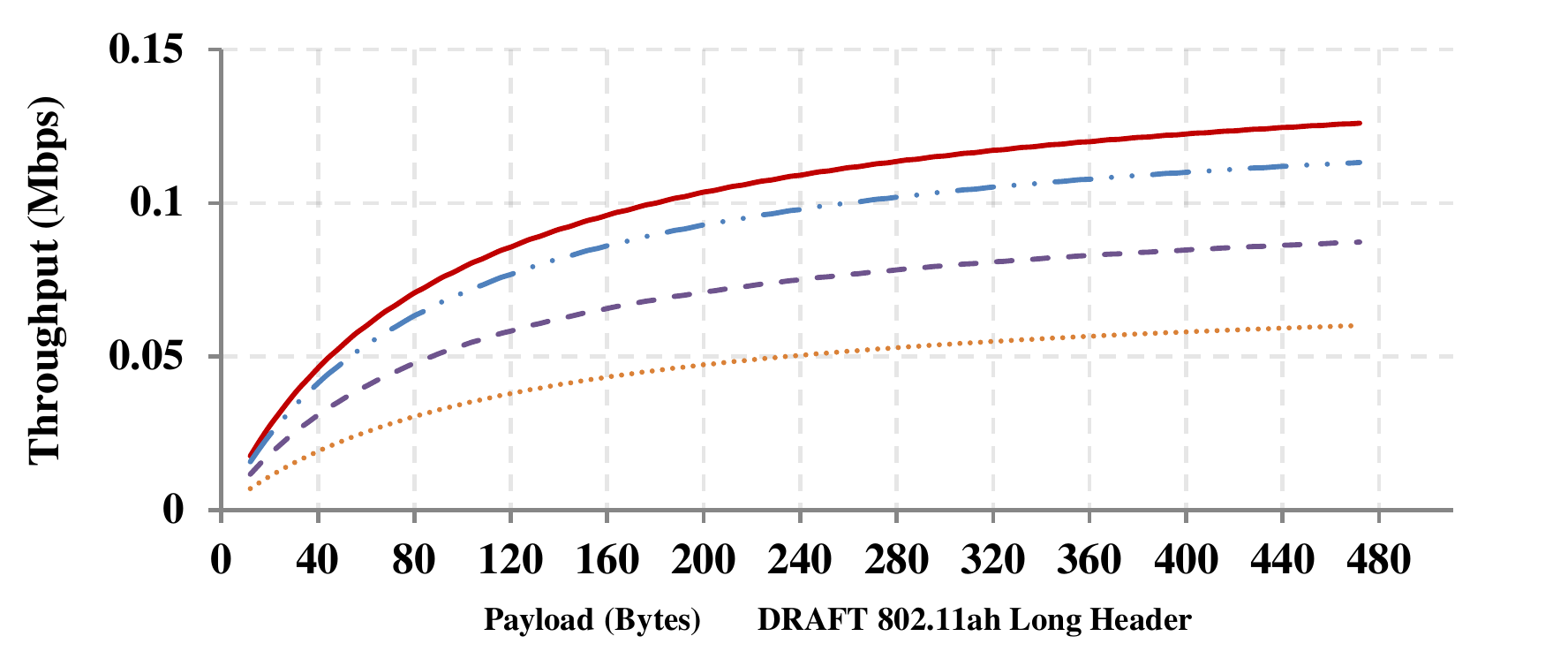}\\
\caption{Throughput vs. Payload size for IEEE 802.11ah (long MAC header and normal ACK).}
\label{fig:fig4}
\vspace{-1.5em}
\end{figure}
\begin{figure}[h]
\centering
\graphicspath{/AHlatex/Figures}
\includegraphics[width=0.46\textwidth]{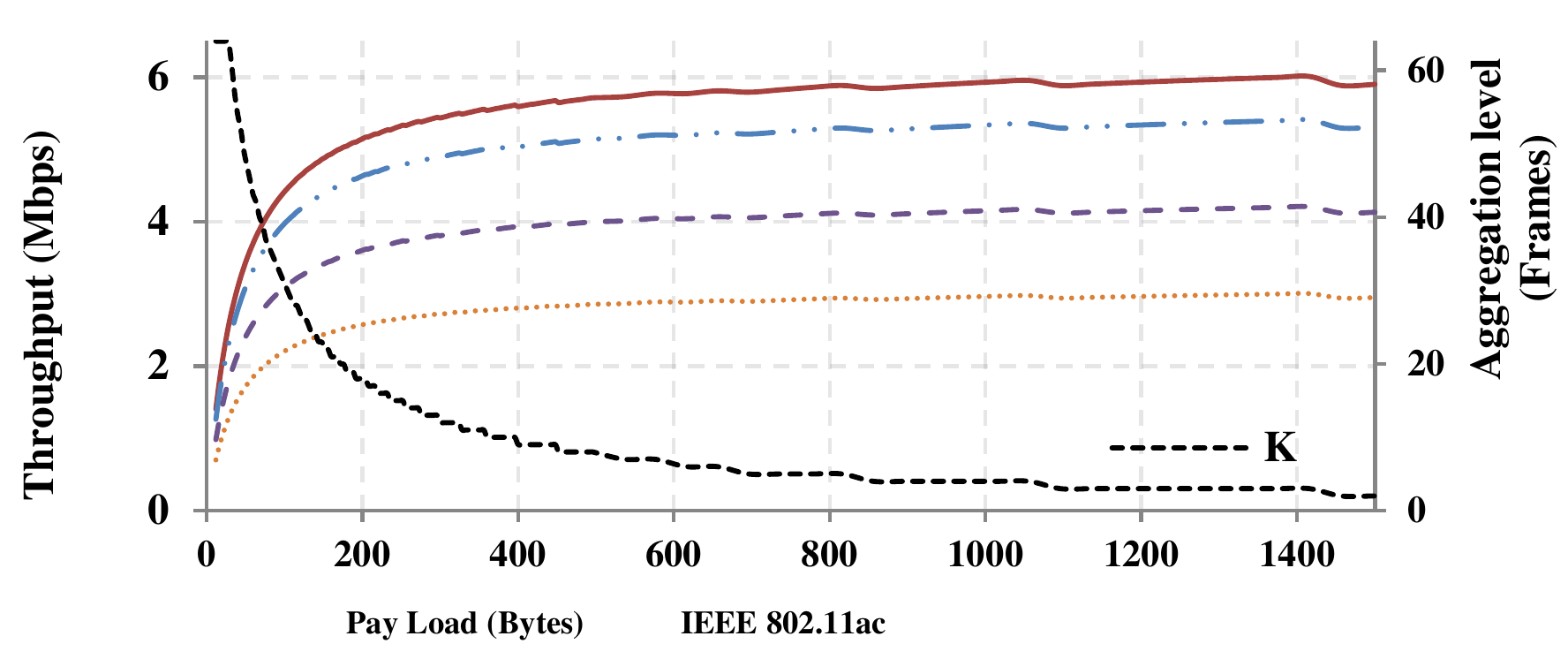}\\
\caption{A-MPDU Throughput vs. Payload size for IEEE 802.11ac.}
\label{fig:fig5}
\end{figure}
\begin{figure}[h]
\vspace{-4em}
\centering
\graphicspath{/AHlatex/Figures}
\includegraphics[width=0.46\textwidth]{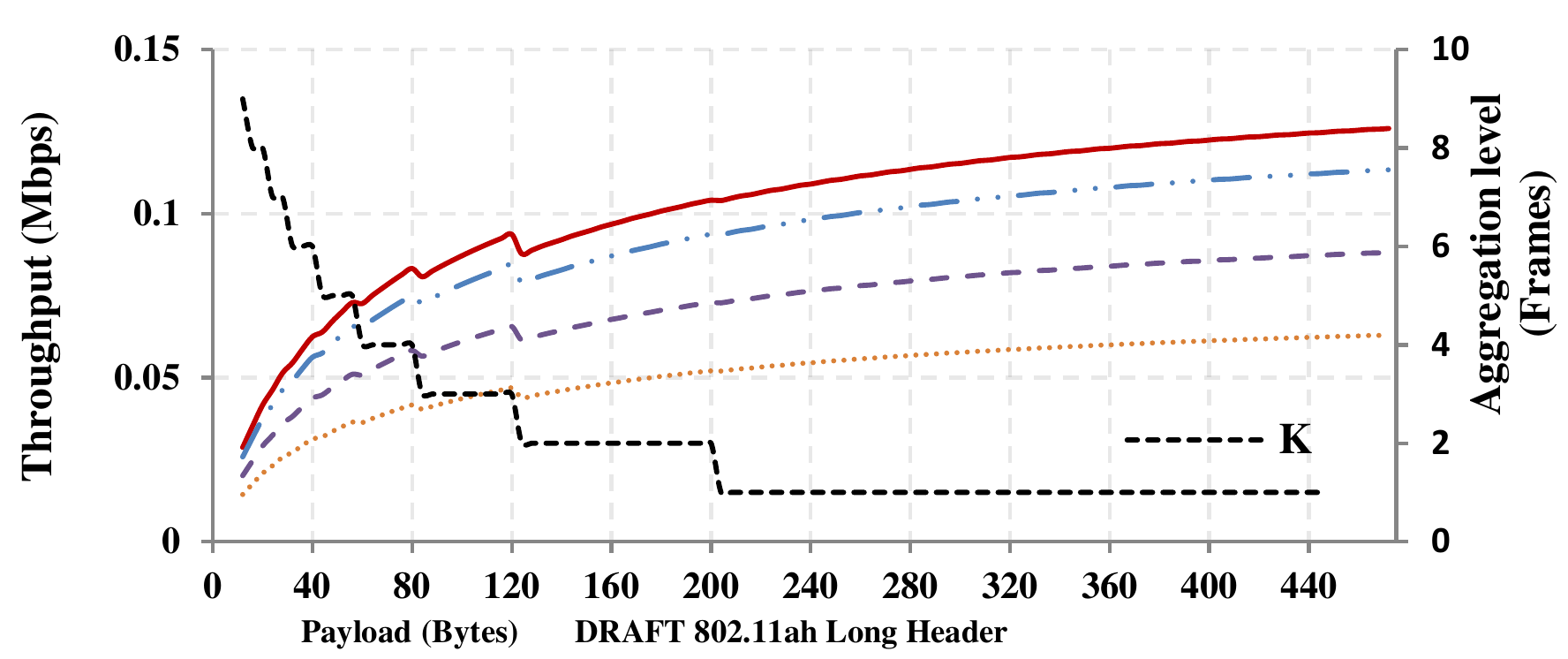}\\
\caption{A-MPDU Throughput vs. Payload size for IEEE 802.11ah.}
\label{fig:fig6}
\vspace{-2.5em}
\end{figure}
\vspace{-.1em}
That is why, in Figure 4, we plot the throughput results using long headers and normal ACK. It is clear that IEEE 802.11ah's most reliable link is much slower than the other technologies (150kbps). Some of the characteristics that give 802.11ah the longest range turn into a drawback when throughput is compared. Results denote a maximum throughput in 475Byte-payload of 126kbps (reduced to 60kbps when \textit{PER} = 50\%). With 12Bytes, throughput varies from 17.6kbps (no errors) to 6.9kbps (\textit{PER} = 50\%).
 In Figure 5, the A-MPDU frame aggregation (IEEE 802.11ac and 11n) provides a maximum throughput (\textit{PER} = 0\%) of 6.71Mbps with 1500Bytes (K=3). \textit{PER} of 50\% reduces to 3.35Mbps. With the shortest frame tested, i.e. 12Bytes, we obtain 1.56Mbps of throughput with 64 MPDUs (K=64) at 0\% \textit{PER} and 0.78Mbps at 50\% \textit{PER} (x4, and x8 times the throughput without aggregation). Considering 11ah and A-MPDU (Figure 6) with the minimum payload (12Bytes of data), we achieve 36.9kbps with 9 MPDUs (K=9) and \textit{PER} equals to 0\%. Note that using the maximum payload, i.e. 475Bytes, there is no aggregation possible. Also note that throughput vs. payload size lines break every time the number of aggregated frames is reduced in order not to exceed the maximum allowed physical frame \cite{stdah} (i.e. less frames can be aggregated as we increase their size).

\section{Conclusion}
In this paper, we provide a comparison between IEEE 802.11ah and current IEEE 802.11a/n/ac in terms of range and throughput. Detailed results indicate that IEEE 802.11ah benefits from the widest coverage, due to the lower frequency band, a narrower bandwidth and a more robust coding scheme. It shows, at least, a fivefold increase with respect to the second best range (IEEE 802.11a), and more than ten times improvement with regard to the worst case (IEEE 802.11n in the 2.4GHz band). On the other hand, IEEE 802.11ah presents the lowest throughput, in comparison with other amendments; a maximum throughput of 144kbps (reduced to 71kbps when \textit{PER} = 50\%) can be achieved at the limits of its estimated coverage, which is enough for the use cases this technology is targeting. Given the use case scenarios where the size of data frames is inherently small, the use of aggregation slightly mitigates the excess of overhead but its impact is limited by the low data rates that 802.11ah devices will support. In our future work, we plan to extend our comparison by means of simulations of more complex scenarios.



\bibliographystyle{IEEEtran}
%
\bibliography{Throughput80211ah}

\end{document}